\documentclass[11pt]{article}
\usepackage{vmargin}
 \setpapersize{A4}
 \setmargrb{.9in}{.78in}{.9in}{.7in}
\usepackage{makeidx}
\usepackage{amssymb}
\usepackage{amsmath, amsthm}
\usepackage{amsfonts, enumerate}
\usepackage{multirow}
\usepackage{color}
\usepackage{caption}

\newcommand{\comment}[1]{}

\begin{document}

\title{\vspace*{-.3in} GPU-Accelerated BWT Construction for Large Collection \\ of Short Reads}

\date{}

\author{
Chi-Man Liu
\hspace*{.5in} 
Ruibang Luo
\hspace*{.5in} 
Tak-Wah Lam
\\[1ex]
\small HKU-BGI Bioinformatics Algorithms and Core Technology Research Center \\
\small Department of Computer Science, University of Hong Kong\\
\small     {\tt \{cmliu, rbluo, twlam\}@cs.hku.hk}
}

\maketitle

\begin{abstract}

Advances in DNA sequencing technology have stimulated the development of algorithms and tools for processing very large collections of short strings (reads).  Short-read alignment and assembly are among the most well-studied problems. Many state-of-the-art aligners, at their core, have used the Burrows-Wheeler transform (BWT) as a main-memory index of a reference genome (typical example, NCBI human genome).  Recently, BWT has also found its use in string-graph assembly, for indexing the reads (i.e.,  raw data from DNA sequencers).  In a typical data set, the volume of reads is tens of times of the sequenced genome and  can be  up to 100 Gigabases. Note that  a reference genome is relatively stable and computing the index is not a frequent task. For reads, the index has to computed from scratch for each given input.  The ability of efficient BWT construction becomes a much bigger concern than before.

In this paper, we present a practical method called CX1 for constructing the BWT of very large string collections. CX1 is the first tool that can take advantage of the parallelism given by a graphics processing unit (GPU, a relative cheap device providing a thousand or more primitive cores), as well as simultaneously the parallelism from a multi-core  CPU and more interestingly, from a cluster of GPU-enabled nodes.  Using CX1, the BWT of a short-read collection of up to 100 Gigabases can be constructed in less than 2 hours using a machine equipped with a quad-core CPU and a GPU, or in about 43 minutes using a cluster with 4 such machines (the speed up is almost linear after excluding the first 16 minutes for loading the reads from the hard disk). The previously fastest tool BRC is measured to take  12 hours to process 100 Gigabases on one machine; it is non-trivial how BRC can be parallelized to take advantage a cluster of machines, let alone GPUs.

With the availability of GPU-enabled clusters like Amazon-EC2, NIH-Biowulf and Tianhe-1A, we believe that our method has given a cost-efficient solution for the bioinformatics community to index a large collection of short reads with BWT.

\end{abstract}

\section{Introduction}

We consider the problem of computing the Burrows-Wheeler transform (BWT) of a large collection of
short strings using a graphics processing unit (GPU), which can be considered  as a low-cost
parallel computing device providing a thousand or more cores
running in the SIMT (single-instruction-multiple-thread) mode.
 The BWT is a string transformation
that has been used widely in indexing texts, such as the human genome (which has 3 billion bases or
characters).  The breakthrough in sequencing technologies has made it feasible to decode a human genome,
at the cost of a few thousand US dollars (soon less than a thousand),
in the form of hundreds of millions to billions of randomly-ordered short fragments, each called a short read
containing
one to several hundred  bases.
 BWT-based indices,
such as the FM-index~\cite{fmindex}, are the
core of many state-of-the-art software tools for DNA mapping short reads onto a known reference genome, for example,
BWA~\cite{bwa}, and Bowtie2~\cite{bowtie2},  SOAP3-DP~\cite{soap3dp}.
These short-read aligners stored the BWT of the reference genome (typical example, NCBI human
genome) in the main memory and the highly optimized operations on BWT indices
allow very fast mapping of individual short reads. Nowadays, aligning a
billion short reads to the human genome only takes a couple of hours on a GPU-assisted computer.

While these aligners are highly optimized on using the BWT indices,
they are not particularly efficient at constructing BWT indices. The reason is that the indices
are built on common reference genomes, which are quite static over time. These indices can be
constructed once and used for a few months, hence not much effort has been put in speeding up the
construction process. It usually takes less than an hour to index a genome of a few Gigabases.
In a typical  sequencing run, the reads
have a told volume tens of times of the reference genome
(for human genome, it is common to have 30+ folds of coverage, translating into
about one billion length-100 reads, with total size about 100 Gigabases).

Indexing the reads is obviously more resource demanding than indexing a reference genome.
Apparently it makes no sense to index the reads.  Only
recently, BWT indexing of short reads is found to be critical and useful for some interesting bioinformatics applications.
The most notable one is the string-graph approach to de novo assembly of short reads to recover the original genome.\footnote{De novo assembly refers to assembly without reference to any reference genome, even if it exists.}
Popular  assembly software are mainly based on de Bruijn graphs (e.g., \cite{soapdenovo2}).   A more traditional but less-used approach is based on string graphs.  Although string graphs are lossless representation of reads and believed to be more useful,
the prohibitive computational complexity of string graph construction makes it infeasible to build a
practical assembler.   Recently, Simpson and Durbin~\cite{sga} observed that
a  string graph can be built very efficiently given the
BWT of  reads to be assembled.
At least two string-graph assemblers, SGA~\cite{sga} and
Fermi~\cite{fermi}, have been successfully developed using this approach.
However, the bottleneck has shifted to the BWT construction, which accounts for more than half of the
overall running time.   
A very efficient  BWT construction  is crucial to make string-graph assemblers practical.
Another application of short-read indices is error correction of short reads based on k-mer counting~\cite{kmer}.
The advantage of BWT is that it supports $k$-mer for different $k$ and allows more flexible and effective error correction.

With the advent in general-purpose GPU computing, the GPU has found to be a cost-efficient
solution to various problems not only in computer graphics, but also in areas like bioinformatics and
computational physics. Several BWT-based short-read aligners are GPU-accelerated, including SOAP3-DP~\cite{soap3dp},
CUSHAW~\cite{cushaw}, and BarraCUDA~\cite{barracuda}, but none of them makes use of the GPU to speed up the construction the BWT of the reference genome (as  it is not time critical).
Recently, Zhao et al~\cite{gpubwt} proposed a method for speeding up BWT construction with the help of GPUs.
Their method outperforms all existing CPU-only BWT construction methods, achieving a 5-fold speedup
 over the CPU-based algorithm by
Lam et al~\cite{bwtinc}. However, GPU has relatively small memory, typically in the range of 2 to 6 Gigabytes.
The largest input Zhao et al's method can handle is about 3~Gigabases.
This is good for indexing a human genome but far from the typical scale of index reads.
Two GPU-accelerated methods for suffix array construction~\cite{ppopp13,spire12} can also be used
to construct the BWT, achieving a speedup of more than 10-fold over similar CPU-only approaches.
However, these methods are again limited by the size of the GPU memory, and can only handle around 200~Megabases.

There are many challenges we need to face when designing a method for constructing the BWT of
a large collection of short reads. The first one is undoubtedly memory. Many existing BWT
construction methods compute an intermediate data structure known as the suffix array. For a string of
100~gigabases, storing its suffix array would need 800~GB of memory, hence we cannot directly adopt
those methods for solving our problem. To make things worse, the GPU can only manipulate data inside
its own device memory, which is no more than 6~GB for modern commercially available GPUs.
In order to harness the power of GPUs,
we must find a way to chop our data into pieces that fit in the GPU memory.

Another challenge is parallelism and scalability, as clusters and multi-core architectures are becoming
ubiquitous.
Most traditional methods for BWT (or suffix array) construction are sequential. To achieve
parallelism, both SGA and Fermi take the approach of partitioning the string collection, constructing
separately the BWT for each partition, and merging the partial BWTs together to get the complete BWT.
However, the final merging step is prohibitively expensive in terms of both computation and I/O.
On the other hand,
GPU's single-instruction-multiple-thread (SIMT) model of execution is an
even bigger obstacle to overcome, and often requires non-trivial optimization.
To parallelize over a cluster of GPU-enabled nodes is even more complicated, yet it
is worth consideration as GPU-enabled cluster service
is getting popular at very reasonable cost (e.g., Amazon-EC2, Norvette, NIH-Biowulf and Tianhe-1A).

Among the existing methods for constructing the BWT of large string collections, BCR~\cite{bcr} (along
with its external memory version) is the most remarkable. With a small memory footprint, BCR constructs
the BWT of 1 billion length-100 reads in half a day. Other approaches taken by SGA and FERMI either take
days to accomplish the same task, or need hundreds of gigabytes of memory. However, BCR has several limits.
Firstly, the method is I/O intensive. During construction, the intermediate BWT stored on disk is
being read and written repeatedly. In fact, as pointed out in their paper, the CPU efficiency of BCR
can be as slow as 20\% for large inputs. Secondly, BCR by default does not utilize all the available
memory. Typical server machines nowadays, especially for bioinformatics applications,
 have 64~GB or more main memory\footnote{Nowadays servers with a few hundred GB of main memory can be available at less than ten thousand US dollars.}, it would be a waste of resources
to limit our memory consumption to, say, 16~GB. Last but perhaps the most critical, it is not obvious how to effectively parallelize
the BCR algorithm using multiple nodes. The BCR algorithm is sequential in nature, that
computing the ``partial BWT'' in each iteration depends heavily on the ``partial BWT'' from the previous
iteration. This limits the scalability of the method severely.

In this paper, we propose a GPU-accelerated method, called CX1, for BWT construction of a large collection of short reads,
which addresses the above issues. Our method tries to minimize I/O operations by utilizing available
memory as much as possible. Multi-threaded CPU and GPU executions are carefully pipelined to maximize the throughput.
Moreover, CX1 can run concurrently on a cluster of nodes to achieve a speedup almost linear
in the number of nodes. Experiments show
that our method can construct the BWT for 1 billion length-100 reads in 2 hours using a quad-core
machine with 64~GB of memory and a 4-GB GPU card, and in around 43~minutes on a cluster of
4 such machines.   
Notice that about 16 minutes of time is spent of loading the reads from the hard disk.  Excluding
the loading time, the speed up on BWT construction is almost linear.

The rest of the paper is organized as follows. Section~\ref{section:prelim} reviews and extends the definition
of the BWT. Section~\ref{section:method} gives the details of our method.
 Section~\ref{section:experiments}
reports the results of our experiments in which our method was evaluated in different aspects such as
speed, memory consumption and scalability. We discuss possible directions for future work
 in Section~\ref{section:conclude}.

\section{Preliminaries}
\label{section:prelim}

In this section, we extend the definition of the Burrows-Wheeler
transform (BWT) for a single string to one for a collection of strings.
We refer readers to~\cite{bwt94} for a detailed discussion
of the BWT. For easier manipulation, all string indices and array indices are 0-based
unless otherwise stated.

Let $\Sigma =  \{ c_0, \ldots , c_{\sigma-1} \}$ be a finite alphabet whose symbols
satisfy $c_0 < \cdots < c_{\sigma-1}$, where $<$ denotes lexicographic order.
Given a string $S = s_0s_1 \cdots s_{m-1}$ of length $m$ over $\Sigma$, we denote
by $S[i,j]$ the substring $s_is_{i+1}\cdots s_j$, for $0 \le i \le j < m$.
A substring $S[i, m-1]$ is called a suffix of $S$. In the context of the BWT, it is
customary to append a special end-marker symbol $\$$ at the end of the string, where
$\$$ is lexicographically smaller than every symbol in $\Sigma$.

Let $R = \{ S_0, \ldots, S_{n-1} \}$ be a collection of $n$ strings. Each of the strings
$S_i$ is of length $m+1$ with $S_i[k] \in \Sigma$ (for $0 \le k < m$) and $S_i[m] = \$$.
For tie-breaking purposes, we assume that the $\$$ symbols associated to different
strings are different, namely, we define $S_i[m] < S_j[m]$ if $i < j$.
To define the BWT of $R$, we first sort lexicographically
the suffixes of all strings in $R$. Each string has $m+1$ suffixes, thus the sorted
list contains $n(m+1)$ different suffixes. Having obtained the sorted list $L$,
we can define the BWT of $R$, denoted by $B_R$, to be a
string of length $n(m+1)$, such that $B_R[i]$ is the symbol preceding the suffix $L[i]$
(in the string of which $L[i]$ is a suffix), for $0 \le i < n(m+1)$. For a suffix
$L[i]$ starting at position 0 of the string, we define its preceding symbol to be $\$$.

Figure~\ref{bwt-example} shows the BWT for the collection $\{ ACGT\$, \, TAGT\$, \, GGAA\$ \}$.

\begin{figure}[h]
\centering
\begin{tabular}[c] { c c  c  c c  c  c  c  c  c  c  c  c  c  c  c }
bwt $\rightarrow$ &  T & T  & A  &  A & G &\$ & T & A & G &\$ & C & A & G & G &\$ \\
\hline
& \$ & \$ & \$ &  A & A & A & A & C & G & G & G & G & T & T & T \\
&    &    &    & \$ & A & C & G & G & A & G & T & T &\$ &\$ & A \\
&    &    &    &    &\$ & G & T & T & A & A &\$ &\$ &   &   & G \\
&    &    &    &    &   & T &\$ &\$ &\$ & A &   &   &   &   & T \\
&    &    &    &    &   &\$ &   &   &   &\$ &   &   &   &   &\$
\end{tabular}
\captionsetup{width=14.5cm}
\caption{ The first row gives the BWT for the collection $\{ ACGT\$, \, TAGT\$, \, GGAA\$ \}$.
All 15 suffixes (under the horizontal line) are sorted lexicographically. The character
above the line precedes the suffix underneath.}
\label{bwt-example}
\end{figure}

\section{Method - CX1}
\label{section:method}

Our approach to constructing the BWT of $R$ follows directly from definition: sorting
suffixes in $R$. Although this sounds easy, directly sorting all suffixes in the GPU is impractical
due to limited memory, even for data sets of moderate sizes.
Consider a set of 100~million strings, each of length 100.
If we explicitly list out all suffixes, there will be roughly 500~billion characters.
The memory required is 125~GB even if we use the standard 2-bit-per-character
representation for DNA strings. This clearly does not fit into the global memory of
any modern GPU. To cope with the tight memory limit, we adopt a strategy similar to blockwise
suffix sorting described in~\cite{Karkka}. In the following, we give the high level idea
of our approach.

Firstly, we find a set of {\em splitters} $\{ P_0, P_1, \ldots, P_u \}$, which are strings over
$\Sigma$ satisfying $P_0 < P_1 < \cdots < P_u$ (they are not necessarily suffixes in $R$).
The suffixes in $R$ are then partitioned by
the splitters, i.e., the first partition contains all suffixes less than $P_0$, the second
partition contains all suffixes in $[ P_0, P_1 )$, and so on, and the last partition
contains all suffixes not less than $P_u$. The splitters are chosen such a way that the suffixes
in a single partition can be sorted within the GPU memory limit. Each partition gives a portion
of the BWT, which can be simply concatenated together to form the complete BWT.
Sections~\ref{section:splitters} and~\ref{section:suff-sort}
explain how to find the splitters and perform suffix sorting with the GPU.
Section~\ref{section:parallel} discusses parallelization of our method.

\subsection{Choosing the splitters and partitioning}
\label{section:splitters}

In the original blockwise suffix sorting algorithm~\cite{Karkka}, the splitters are chosen
by randomly sampling the suffixes in $R$ and sorting them.
The approach we take is as simple --- the splitters are
the set of strings over $\Sigma$ of the same length, say $\ell$. Intuitively, suffixes with
the same $\ell$-prefix (i.e. the first $\ell$ characters) are in the same partition (suffixes
shorter than $\ell$ require special but easy handling).

To determine the partition to which each suffix belongs, we simply scan all suffixes and examine
their $\ell$-prefixes. Recall that the reason for partitioning suffixes is to let each
partition fit into the GPU. If it happens that some partition is still
too large to fit into the GPU, we further partition it by looking at the next $\ell$
(or fewer, depending on the size of the partition) characters of the suffixes in that partition.
This can be done
recursively until every partition can fit into the GPU. On the other hand, small partitions are
not favorable for the GPU. To maximize throughput for the GPU, we would like each partition to be
as close to the GPU memory limit as possible. This is achieved by merging a consecutive
range of partitions such that the total size is just about to exceed the
GPU memory limit.

So far we are not clear about how to list all suffixes in every partition in main memory.
Storing all suffixes
explicitly takes $O(nm^2 \log \sigma)$ bits of space, or 125~GB for our moderate-sized example
at the beginning of
Section~\ref{section:method}. A large data set could occupy well over 1~TB of memory. Clearly,
we need a more compact way to represent the partitions.
We use the natural solution of specifying a suffix by its starting position in the string
collection $R$, viewing $R$ a string formed by concatenating the enclosed strings in order.
This approach needs $O(nm)$ words of space (assuming the value $nm$ fits into a word).
However, for large data sets with 1 billion reads, this still amounts
to over 100~GB of memory. In case this does not fit into the main memory of the machine,
we need to perform partitioning in multiple rounds. The idea is that in each round, we only
only list the suffixes in some selected partitions, sort those suffixes, output their
associated BWT characters, and release the suffixes from main memory for the next round.

Figure~\ref{algo} gives the outline of this algorithm.
First we need two thresholds $m_1$ and $m_2$ ($m_1 \le m_2$). $m_1$ represents the maximum number
of suffixes that
can be sorted by the GPU at one time, which will be discussed in Section~\ref{section:suff-sort}.
$m_2$ represents the maximum number of suffixes (stored as positions in $R$) that can be stored
in the available main memory. Initially, we precompute the size of every partition by scanning
all suffixes as discussed above, and store the sizes in the $psize$ array. Then, in each round,
we select a range of partitions with total size not exceeding $m_2$, and list out the suffixes
in the selected partitions (as starting positions). Since $m_2 \ge m_1$, most likely these
suffixes will not all fit into the GPU. So we need to sort them in {\em small rounds}, where
in each small round we explicitly list out at most $m_1$ suffixes in a subset of the
selected partitions, and have them sorted with the GPU (Section~\ref{section:suff-sort}).
After all suffixes  in the selected
partitions have been sorted, we discard them from main memory and proceed to the next round.

The parameter $m_2$ can be adjusted to control main memory consumption. A larger $m_2$ would
mean fewer number of rounds (and thus fewer scans of the suffixes), but at the expense of
a larger memory footprint.

\subsection{GPU suffix sorting}
\label{section:suff-sort}

At the core of our BWT construction method is suffix sorting with the GPU. The GPU excels at
sorting numbers or records with numeric keys; for example, the Thrust library provides
GPU implementations for merge sort and radix sort. The idea we use for sorting strings is
based on radix sort. Suppose that all the strings we want to sort have the same length,
say $k$ (we will remove this assumption later).
Each string can be viewed as a $\sigma$-ary integer with $k$ digits, so
least-significant-digit (LSD) radix sort can be applied to sort the strings in $k$
iterations. To reduce the number of iterations, we can sort multiple digits in a single
iteration. For example, a DNA string in 2-bit-per-character representation can be sorted
by 16 bits (i.e. 8 characters) per iteration.

Recall that radix sort permutes all elements in every iteration. Strings are really big
elements spanning multiple words, thus swapping them in memory takes time. A typical solution
is sorting by index. We assign an integer index to each string before sorting. When sorting,
instead of swapping the strings, we swap the indices which are relatively small. The
downside of sorting by index is that we need an extra memory redirection to access the string
via the index. However, this redirection overhead is compensated by the savings in swapping.

We can take this idea a step further to minimize GPU memory usage.
In order to let GPU do the sorting, we first need to copy the strings and the indices to
the GPU memory. The longer the strings, the fewer of them can reside in the GPU, thus
lowering the throughput of sorting. To allow more strings to be sorted by the GPU
at one time, we observe that radix sort only looks at a certain number of bits of each string
in an iteration. Therefore, we only need to keep the indices and the required bits of the strings
in the GPU for an iteration. As an example, consider sorting DNA strings of length 64. Each
string is represented as 128 bits, with an associated 32-bit integer index. For brevity,
we view a string as four 32-bit words called {\em stringwords}.
Suppose that we
use radix sort on key-value pairs with 32-bit keys and values. In the first iteration,
we copy the index and the lowest stringword of every string into the GPU. Then, we perform radix
sort with the bits as keys and the indices as values. After sorting by the keys (stringwords),
the indices reflect the ordering of the strings by the lowest stringword. In the next iteration,
we copy the second-lowest stringword of every string into the GPU, replacing the lowest stringwords.
The new stringwords do not match the indices already in the GPU, so we need to
permute the stringwords according to the indices, which is trivial for the GPU. Then, we perform
key-value radix sort as in the first iteration. After four iterations, the index array in the GPU
gives the sorted order of the strings. With this approach, we need only $32+32=64$ bits of GPU memory
for each string, instead of $128+32 = 160$ bits. From this discussion, we see that the maximum
number of suffixes that can be sorted with the
GPU at one time, i.e. $m_1$, is not related to the length of the suffixes, but only depends on
the GPU memory limit (and word size).

Before we end this section, we describe how to remove the assumption that all strings to be
sorted have the same length. Recall that we are sorting suffixes which all end with the $\$$ symbol.
For every suffix shorter than $m+1$, we append $\$$ to it to make its length $m+1$. (This resembles
adding leading zeroes to integers in radix sort.) Another requirement we stated earlier is that
the $\$$ symbols associated with different strings should be considered different, namely,
the $\$$ in $S_i$ is lexicographically smaller than that in $S_j$ if $i < j$. This is easily fulfilled
by listing suffixes in a partition so that suffixes of $S_i$ come before those of $S_j$ if $i<j$,
and using stable radix sort to preserve the orders of equal suffixes.

\subsection{Parallelization}
\label{section:parallel}

Our method can be parallelized at two levels. At the high level, we can make
it run on multiple nodes to achieve speedup almost
linear in the number of nodes. Suppose that we have $N$ nodes, each with a copy of the input data set
on disk. We dedicate one of the them as the master node, which scan all suffixes
and computes the $size$ array (Step 1). It then divides
the set of all partitions into $N$ ranges such that the total partition sizes of the ranges are
roughly equal. Each node is responsible for sorting partitions in one of the $N$ ranges, as
assigned by the master node. After every node has finished computing its portion of the BWT,
the pieces can be sent back to the master node for direct merging to get the complete BWT.
The only sequential steps are computing $size$ and merging at the end.

At the low level, almost every computationally intensive step in our method can be made to run
on parallel threads. For the steps which requiring scanning all suffixes, each thread may
be responsible for the suffixes of a subset of strings. For instance, if we have two threads,
the first thread would scan all suffixes of $\{ S_0, \ldots, S_i \}$, while the second thread
would scan all suffixes of $\{ S_{i+1}, \ldots, S_{n-1} \}$, where $i = \lfloor n/2 \rfloor$.
Each thread has its own $size$ array, which can be merged together to form the global $size$ array.

\begin{figure}[t]
\centering
\framebox[0.9\textwidth][c]{
\parbox[c][][c]{0.8\textwidth}{
{\bf Input:} A string collection $R$. \\
{\bf Parameters:} $\ell$ is the prefix length for partitioning.
$m_1$ and $m_2$  are two thresholds depending on the memory limits.
$m_1$ scales with the GPU global memory size, and $m_2$ scales with the main memory size. \\
{\bf Output:} The BWT of $R$.
\begin{enumerate}
\item Scan all suffixes of $R$.
For each suffix, examine its $\ell$-prefix to determine its partition.
Store the size of each partition in an array $size$.
If the size of a partition exceeds $m_1$, refine it by examining a
longer prefix.
\item Divide the set of all partitions into
ranges, such that the total partition size of each range is at most $m_2$.
\item For each range $[a, b]$ do
\begin{enumerate}
\item  Scan all suffixes and record the positions of those whose partition is in $[a, b]$.
The positions should be sorted according to the partitions they are in.
\item Further divide $[a, b]$ into {\rm small ranges}, such that the total partition size of each small range
is at most $m_1$.
\item For each small range $[a', b']$ do
\begin{enumerate}
 \item For each suffix position whose partition is in $[a', b']$, copy that suffix to an array $W$.
 \item Sort the suffixes in $W$ with the GPU.
 \item For each suffix in the sorted order, output its preceding character as the next BWT character.
\end{enumerate}
\end{enumerate}
\end{enumerate}
}
}
\captionsetup{width=14.5cm}
\caption{ Outline of our BWT construction algorithm CX1.}
\label{algo}
\end{figure}

\section{Experiments}
\label{section:experiments}

We implemented our BWT construction method CX1  in C++ and CUDA. For GPU radix sort, we used the library
by Merrill and
Grimshaw~\cite{radix}~\footnote{The radix sort library is available at https://code.google.com/p/back40computing/wiki/RadixSorting}.
We tested our implementation
with the Asian YH short reads in~\cite{soapdenovo2}. The results were compared
to BCR (version 0.6.0)~\footnote{BCR is part of the BEETL library available at
http://beetl.github.io/BEETL/}. All single-machine tests were carried out on a machine with
a Intel Xeon i7-3930K (Hexa-core) 3.2~GHz processor and 64~GB of memory. The machine
is also equipped with an NVIDIA GeForce GTX 680 graphics card with 4~GB of global memory.
Tests on multiple nodes were carried out on a cluster of 4 such machines connected by
Gigabit Ethernet.

We prepared three input data sets with 10~million, 100~million, and 1~billion reads respectively.
For each data set, we arbitrarily selected a subset of the YH short
reads of the required size. Each read was 100 bases long and consisted only of $\{ A, C, G, T \}$.
The short reads were then passed as input to our program and BCR in FASTA format. The BWT
was outputted in ASCII format. All parameters and options of BCR except ``I/O formats'' and
''algorithm'' (which was set to ``BCR'') were left as default.

\subsection{Construction time and memory}

Table~\ref{exp-time} shows the construction time and memory consumption for CX1 and BCR on
three data sets when running on a  computer equipped with a quad-core CPU and a GPU.
Note that BCR can only utilize one CPU core.   
For the data set with 1,000 M reads (total volume 100 Gigabases), CX1 takes 6,886 seconds and BCR
46,899 seconds.
CX1 is measured to have higher memory utilization than BCR, especially for smaller data sets.
This is because CX1 is designed to take advantage of all available memory.
For algorithmic interest, CX1's memory usage can be limited by setting a smaller value for
the parameter $m_2$.  E.g., for the 100 M data set,  CX1 can reduce the memory usage
from 45 GB to 16 GB, and the time slightly increase from 1,898 seconds to 2,024 seconds.

It is worth-mentioning that our experiment was done on a quad-core CPU, which is rather outdated when compared to
the servers nowadays (6 to 16 cores).
  CX1 can be adjusted to use more cores, yet it is unlikely to be effective for small data sets as
the bottleneck resides in the GPU.
For large data sets, the bottleneck of CX1 gradually shifts to the main memory access
and the CPU, and CX1 will perform better with more cores in the CPU.  

\begin{table}[t]
\centering
\begin{tabular}{c|c c c}
\hline
     & 100M reads & 500M reads & 1000M  reads\\
\hline
 BCR &  6,141    &    23,094 &   46,899 \\
     &  3.3 GB  &   17.6 GB &   33.3 GB \\
\hline
 CX1 & 565 & 3,108 & 6,886 \\
    & 45.0 GB &  57.0 GB & 57.0 GB \\
\hline
\end{tabular}
\captionsetup{width=14.5cm}
\caption{ Construction time (in seconds) and memory consumption for three data sets using
one machine (with a GPU).
All reads are of length 100.}
\label{exp-time}
\end{table}

\begin{table}
\centering
\begin{tabular}[c] { c | c c c }
\hline
   \# machines  & 100M reads & 500M reads & 1000M reads \\
\hline
 1 &   565  &    3108 &  6886 \\
  &  (1.00) & (1.00) & (1.00) \\
\hline
 2 & 338 & 1797 & 3998 \\
  & (1.67)  & (1.73) & (1.72) \\
\hline
 3 &  266  & 1426 & 3071 \\
  & (2.12) & (2.18) & (2.24) \\
\hline
 4 & 230 & 1192 & 2584 \\
  & (2.46) & (2.61) & (2.66) \\
\hline
\end{tabular}
\captionsetup{width=14.5cm}
\caption{ Construction time of CX1 (in seconds) and speedup (in brackets, relative to a single node)
  using up to 4 GPU-enabled machines. All reads are of length 100.}
\label{exp-node}
\end{table}

\subsection{Cluster parallelism}


We have tested CX1 using a cluster of 2 to 4 nodes (computers), each equipped with a quad-core CPU
and a GPU.  The time required is shown in
Table~\ref{exp-node}. 
For 1,000 M reads, 
CX1 takes 43 minutes using 4 nodes,
and less than 2 hours using 1 node.   The speedup is 2.66, not particularly impressive.  In fact,
the time measured includes more than 16 minutes on loading the reads from the hard disk, which remains
the same using different number of nodes.   To better measure the
speedup on BWT construction,
we exclude the loading time and show the actual construction time in Table~\ref{exp-noio}.
For the data set of 1,000 M reads, the speedup from 1 node to 4 nodes becomes 3.72.
The
speedup is getting close to linear to the number of nodes, especially
for large data set.
We have not tested but believe that similar speedup will be obtained even for a larger cluster with up to ten nodes.
In conclusion, CX1,
given enough resources for parallelism,
can give a very efficient solution for BWT indexing of reads.
Compared to
 the sequential BCR, CX1 can be 20 times faster even using 4 computers in parallel.

\subsection{Effects of read length}

We performed experiments to illustrate the effects of different read lengths on the
construction time. Based on the 100M data set, we created an additional data set with
the same number of total bases, by concatenating two adjacent 100-bp reads into one
200-bp read. Similarly, we created another additional data set by concatenating
four adjacent 100-bp reads into one 400-bp read. The new data sets have
50M and 25M reads respectively. From  Table~\ref{exp-len}, we see that BCR needed 52\% more time
for 200-bp reads, and 290\% more time for 400-bp, while our method needed 28\% and 125\%
more. This shows that our method is much less sensitive to longer reads than BCR does.

\section{Discussion}
\label{section:conclude}
We presented a method for constructing the BWT of a large string collection
with the GPU.
As supported by experiments, our method is highly efficient and scalable compared
to existing tools.

One direction for improvement is optimizing I/O. In our experiments with multiple
nodes, reading the input FASTA file contributed a significant amount of running
time. To further improve scalability, we need to minimize the time for reading
input. On the hardware side, we can use an SSD or multiple disks; on the
software side, we can compress the FASTA file before distributing it among the
nodes. This can also reduce network transfer time.

Our experiment results also point out that for large data sets, the workload on
the CPU side dominates the overall running time, thus adding more GPUs to the
system would not help. It is worthwhile to study how we can shift more workload
to the GPU.

\renewcommand{\tabcolsep}{20pt}
\renewcommand{\arraystretch}{1.1}


\begin{table}[t]
\centering
\begin{tabular}[c] { c | c c c }
\hline
   \# machines  & 100M reads & 500M reads & 1000M  reads\\
\hline
 1 &   468  &    2624 &  5882 \\
  &  (1.00) & (1.00) & (1.00) \\
\hline
 2 & 241 & 1317 & 2994 \\
  & (1.94)  & (1.99) & (1.96) \\
\hline
 3 &  169  & 944  & 2067 \\
  & (2.77) &  (2.78) & (2.85) \\
\hline
 4 & 133 & 712 & 1580 \\
  & (3.52) & (3.69) & (3.72) \\
\hline
\end{tabular}
\captionsetup{width=14.5cm}
\caption{  Figures in Table~\ref{exp-node} revised by excluding the time spent
on loading the reads.}
\label{exp-noio}
\end{table}

\begin{table}
\centering
\begin{tabular}[c] { c | c c c }
\hline
    & 100M (100-bp) & 50M (200-bp) & 25M (400-bp) \\
\hline
 BCR &   6141  &  9334  &  23950 \\
\hline
CX1 & 565 & 724 & 1269 \\
 \hline
\end{tabular}
\captionsetup{width=14.5cm}
\caption{ Construction time 
for different read length;  only one machine is used.}
\label{exp-len}
\end{table}

\end{document}